\documentclass[
showpacs,aps,prb,amsmath,amssymb]{revtex4}
\usepackage{graphicx}
\begin{document}

%
\title{Effective t-J model of pairing: singlet versus triplet}
\author{Jozef Spa\l{}ek}
\affiliation{$^{1}$Marian Smoluchowski Institute of Physics, Jagiellonian
University, Reymonta 4, 30-059 Krak\'ow, Poland \\
e-mail: ufspalek@if.uj.edu.pl \\
URL: http:$//$th-www.if.uj.edu.pl$/$ztms$/$jspalek$\_$een.htm}

\begin{abstract}
The t-J model is regarded as a canonical model of spin-singlet pairing
induced by the {\em kinetic exchange interaction\/} responsible also
for an antiferromagnetic ordering in the strongly correlated narrow-band
systems. In the orbitally degenerate systems both ferromagnetic and
antiferromagnetic kinetic exchange interactions occur. I review briefly
the analogy between the singlet and triplet types of pairing, as well as
draw some general conclusions about the pairing induced by these exchange
interactions. The general discussion is also illustrated with a concrete
case of a two-dimensional lattice with the spin triplet pairing.
\end{abstract}
\maketitle

\section{Introduction}

The t-J model contains a very attractive idea of the spin-singlet pairing
induced by electron correlations and taking place
in the high-temperature superconducting cuprates [1,2], although some
questions remain whether the actual pairing should include
explicitly also the hybrid
$2p-3d$ pairing induced by the interactions of the Kondo-type [3,4]. What
is fundamental in this approach is the crucial role of a strong Coulomb
repulsion between electrons in these narrow-band systems, which leads to the
antiferromagnetic intersite interaction, yielding in turn the effective
pairing in {\em real space\/}. Even the effect of the atomic disorder leading to an
enhancement of the pairing on local scale can be accounted for
semiquantitatively [5]. The question of the pseudogap appearance in the
underdoped systems is still under debate, but it may be related to the
residual antiferromagnetic correlations in the strongly disordered medium,
induced by both the doping and the correlations of charge and spin
type (stripes).
In general, the nature of the normal state is understood to a lesser extent
than the condensed superconducting state in the two spatial dimensions.
The role of the third dimension (coupling between the planes), as well as
that of electron-lattice coupling, remains still at a further distance.

On the other hand, the first unambiguous defection of spin-triplet
superconductivity in the ferromagnetic phase [6,7] did not spark
corresponding modeling of the spin-triplet pairing based on an analogous
approach, as it deserved (see however [8,9]). The systems such as UGe$_{2}$
or URhGe may not be as strongly correlated as the cuprates or heavy fermions,
but still it is important to explore the generalized t-J model for orbitally
degenerate systems. This topic will be overviewed briefly here. In connection
with this one should mention the just discovered Fe-pnictide superconductors,
which are orbitally degenerate systems, so they may be good candidates for a
singlet-triplet superconducting transition as a function of composition
(i.e. concentration of $3d$ electrons). The presence of such a transition would
provide an unambiguous evidence for the role of kinetic exchange interactions
in real-space pairing in orbitally degenerate narrow-band systems.
The purpose of this paper is first to comment of two nonstandard features
of the t-J model and then, to turn attention to an effective t-J model
for orbitally degenerate systems.

\section{t-J model}

The t-J model with inclusion of antiferromagnetic kinetic exchange
and the three-site hopping has been derived over 3 decades ago [9,10].
Starting from the Hubbard model it has the following form for electron
moving in the lower Hubbard band
$$
P_{0}\tilde{H}P_{0}=\sum_{ij\sigma} t_{ij}
a_{i\sigma}^{\dagger}(1-n_{i\overline{\sigma}})a_{j\sigma}
(1-n_{j\overline{\sigma}})
+\sum_{ij} {'}(2t_{ij}^{2}/U)
\left[ {\bf S}_{i}\cdot{\bf S}_{j}-\frac{1}{4}
\sum_{\sigma^{\prime}}
n_{i\sigma}(1-n_{i\overline{\sigma}})
n_{j\sigma^{\prime}}(1-
n_{j\overline{\sigma}^{\prime}})\right]
$$
\begin{equation}
+\, \sum_{ijk}\, \frac{t_{ij}\,t_{jk}}{U}\,
\left[ a_{i\sigma}^{\dagger}\, (1\,-\, n_{i\overline{\sigma}})\,
n_{j\overline{\sigma}}\,
(1\,-\, n_{j\sigma})\, a_{k\sigma}\,
(1\,-\, n_{k\overline{\sigma}})
-\, a_{i\sigma}^{\dagger}\,
(1\,-\, n_{i\overline{\sigma}})\, {\bf S}_{j}^{-\sigma}\,
a_{k\overline{\sigma}}\, (1\,-\,n_{k\sigma})\right]\,.
\label{eq:w1}
\end{equation}

The first term represents the hopping between the neighbors $i$ and $j$
($t_{ij}\neq 0$ for $i\neq j$ only),
the second the so-called kinetic exchange interaction
with the fermionic representation of the spins:
${\bf S}_{i}\equiv (S_{i}^{+},S_{i}^{-},S_{i}^{z})\equiv
(a_{i\uparrow}^{\dagger} a_{i\downarrow}, a_{i\downarrow}^{\dagger}\,
a_{i\uparrow}, (n_{i\uparrow}-n_{i\downarrow})/2)$.
The three processes composing the effective Hamiltonian are depicted
schematically in Fig. 1.

\begin{figure}[htb]
\centerline{\includegraphics[width=0.65\textwidth]{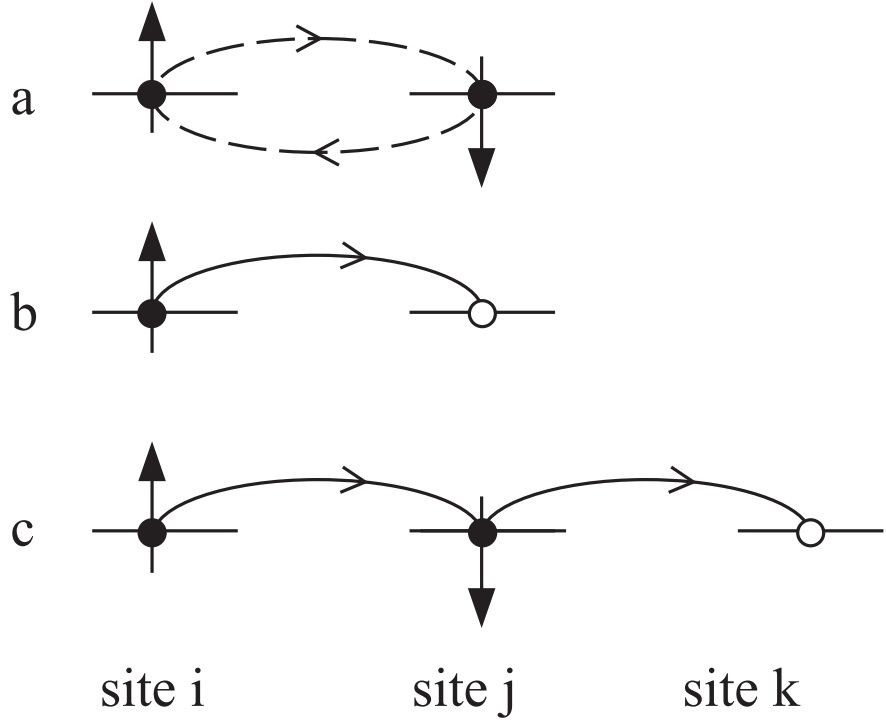}}
\caption{Possible hopping process in the lower Hubbard subband:
a) virtual hoping between singly occupied sites that leads to an
antiferromagnetic (kinetic) exchange interaction between the neighboring
sites; b) two-site hopping between empty and singly occupied sites
(single-particle hopping); and c) between the singly occupied and empty
sites via a singly occupied site.}
\label{fig:f1}
\end{figure}

The whole dynamics
can be expressed in the closed form by introducing projected fermionic
operators:
$$
b_{i\sigma}^{\dagger}\,\equiv\, a_{i\sigma}^{\dagger}\,
(1\,-\, n_{i\overline{\sigma}})\,,\,
b_{i\sigma}\,\equiv\, a_{i\sigma}\,
(1\,-\, n_{i\overline{\sigma}})\,,
$$
\begin{equation}
\nu_{i\sigma}\,\equiv\, b_{i\sigma}^{\dagger}\, b_{i\sigma}\,=\,
n_{i\sigma}\, (1\,-\, n_{i\overline{\sigma}})\,,\,
\nu_{i}=\sum_{\sigma} \,\nu_{i\sigma}\,,
\label{eq:w2}
\end{equation}
and then the Hamiltonian takes the following closed form:
$$
P_{0}\, \tilde{H}\, P_{0}\,=\, \sum_{ij\sigma} {'} \, t_{ij}\,
b_{i\sigma}^{\dagger}\, b_{j\sigma}\,
+\,\sum_{ij} {'}\, \left(\frac{2\, t_{ij}^{2}}{U}\right)\,
\left( {\bf S}_{i}\,\cdot\,{\bf S}_{j}\,-\,\frac{1}{4}\,
\nu_{i}\,\nu_{j}\right)
$$
\begin{equation}
+\, \sum_{ijk}\, \frac{t_{ij}\,t_{jk}}{U}\,
\left( b_{i\sigma}^{\dagger}\, \nu_{j\overline{\sigma}}\,
b_{k\sigma}\,-\, b_{i\sigma}^{\dagger}\,S_{j}^{-\sigma}\,
b_{k\overline{\sigma}}\right)\,.
\label{eq:w3}
\end{equation}

As the projected operators $\{ b_{i\sigma}^{\dagger}\}$
and $\{ b_{i\sigma}\}$ obey non-fermion anticommutation relations
from one side and the hopping term does not commute with the second
order $(\sim \, t^{2}/U)$ part, the dynamics of this model is highly
nontrivial. Physically, this is because the kinetic-energy part vanishes
in the limit of the Mott insulator ($<\nu_{i}>=1$).

An important comment can be made at this point. Namely, in the low-dimensional
systems the Coulomb interaction screening is less effective (particularly
for small number of holes in the Mott insulator). Therefore, the intersite
interaction part
$$
\frac{1}{2}\, \sum_{i\neq j}\, K_{ij}\, n_{i}\, n_{j}\,,
$$
can be regarded as important, particularly near the Mott-insulator limit.
In deriving the effective Hamiltonian the coupling
constant for the second-order terms is then $\sim t_{ij}t_{jk}/(U-K_{ij})$,
which means that the effective spin-spin interaction becomes stronger.
This is not so good for the real space pairing, as one has to add to the
effective Hamiltonian (\ref{eq:w3}) the intersite repulsion $(1/2)
\sum K_{ij}\, \nu_{i} \nu_{j}$ and if $K_{ij}>4t_{ij}^{2}/(U-K_{ij})$, then the
spin-spin coupling is not sufficient to produce the attractive interaction,
i.e. to produce
bound pairs in the ordinary sense.
To see this more
explicitly, we introduce the intersite spin-singlet pairing operators as follows
$$
B_{ij}^{\dagger}\,\equiv\,\frac{1}{\sqrt{2}}\,
\left( b_{i\uparrow}^{\dagger}\, b_{j\downarrow}^{\dagger}\,-\,
b_{i\downarrow}^{\dagger}\, b_{j\uparrow}^{\dagger}\right)\,,
$$
\begin{equation}
B_{ij}=(B_{ij}^{\dagger})^{\dagger}\,=\,-\frac{1}{\sqrt{2}}\,
\left( b_{i\uparrow}\, b_{j\downarrow}\,-\,
b_{i\downarrow}\, b_{j\uparrow}\right)\,,
\label{eq:w4}
\end{equation}
and hence (\ref{eq:w3}) takes the most compact form
\begin{equation}
P_{0}\tilde{H}P_{0}\,=\,
\sum_{ij\sigma} {'}\, t_{ij}\,
b_{i\sigma}^{\dagger}\, b_{j\sigma}\,-\,
\sum_{ijk}\, \frac{2\, t_{ij}\, t_{jk}}{U}\,
B_{ij}^{\dagger}\, B_{jk}\,+\,\frac{1}{2}\,\sum_{i\neq j}\,K_{ij}\,
\nu_{i}\,\nu_{j}\,.
\label{eq:w5}
\end{equation}

The destructive role of the intersite Coulomb repulsion is usually
ignored (see also below).
However, taken literally, it simply can preclude any bound-state
formation associated with the real-space pairing, i.e. when $<B_{ij}>\neq 0$.
To overcome this intersite repulsion a local lattice-distortion induced
attraction can play an important role. An elementary argument for the influence
of the distortion on the effective attractive interaction is briefly
analyzed below [10,11].

\section{Bond distortion and the effective attractive interaction in t-J model}

In the narrow-band limit the hopping integral diminishes roughly exponentially
with the increasing inter-atomic distance ${\bf R}_{ij}$. On the contrary,
in this asymptotic limit the intersite long-range
Coulomb repulsion can be approximated
to the first order by its classical expression $K_{ij}\approx e^{2}/R_{ij}$.
Therefore the change of the system energy $\delta H$ under the influence
of the classical bond distortion $\delta {\bf R}_{ij}$ is mainly influenced by the change
of the intersite Coulomb energy via
$(\delta K_{ij}/\delta {\bf R}_{ij})\cdot\delta {\bf R}_{ij}$, i.e. as
\begin{equation}
\frac{\delta K_{ij}}{\delta {\bf R}_{ij}}\,\simeq\,-\frac{e^{2}}{2 R_{ij}^{2}}
\,\frac{{\bf R}_{ij}}{R_{ij}}\,.
\label{eq:w7}
\end{equation}

We can write down the total system energy with inclusion of the lattice
distortion in the form (we omit three-site terms for simplicity):
$$
\tilde{H}\,=\,P_{0}\tilde{H}P_{0}\,+\,\delta H\,=\,
\sum_{ij\sigma}\, t_{ij}\,
b_{i\sigma}^{\dagger}\, b_{j\sigma}\,+\,
\sum_{ij}\, \frac{2\, t_{ij}^{2}}{U-K_{ij}}\,
B_{ij}^{\dagger}\, B_{ij}
\,+\,\frac{1}{2}\,\sum_{i\neq j}\,K_{ij}\,
\nu_{i}\,\nu_{j}
$$
\begin{equation}
-\,\frac{1}{2}\,\sum_{i\neq j}\,W_{ij}\,
\frac{{\bf R}_{ij}\,\cdot\,\delta {\bf R}_{ij}}{R_{ij}^{2}}\,\nu_{i}\,\nu_{j}
+\,\frac{1}{2}\,\sum_{i\neq j}\,\gamma_{1}\,(\delta {\bf R}_{ij})^{2}\,+\,
\gamma_{2}\,\sum_{\alpha\neq \beta}\,
\delta R_{ij}^{\alpha}\,\delta R_{ij}^{\beta}
\,.
\label{eq:w8}
\end{equation}
where $W_{ij}\simeq K_{ij}\simeq
e^{2}/(2R_{ij})$, $\gamma_{1}$ and $\gamma_{2}$
are the diagonal and off-diagonal elastic constants, respectively.
Minimizing the energy with respect to classical bond distortions
$\delta {\bf R}_{ij}$, one obtains [11] the following effective Hamiltonian
including both strong correlations and the attractive interaction induced
by the particle motion (e.g. holes in the Mott insulator). Assuming
that we have an isotropic distortion accompanying the hole motion
in the planar case, we obtain the following generalized t-J Hamiltonian
(up to an irrelevant term):
\begin{equation}
\tilde{H}\,=\,
\sum_{ij\sigma}\, t_{ij}\,
b_{i\sigma}^{\dagger}\, b_{j\sigma}\,
-\,\sum_{ij}\,\left(
\frac{4t_{ij}^{2}}{U-K_{ij}}\,+\,\frac{1}{4}\,\frac{W_{ij}^{2}}{R_{ij}^{2}\gamma_{1}}\,-\,
K_{ij}\right)\,b_{i\uparrow}^{\dagger}\, b_{j\downarrow}^{\dagger}\,
b_{j\downarrow}\, b_{i\uparrow}\,,
\label{eq:w9}
\end{equation}
or equivalently,
\begin{equation}
\tilde{H}\,=\,
\sum_{ij\sigma}\, \left( t_{ij}\,
b_{i\sigma}^{\dagger}\, b_{j\sigma}\,-\,\frac{1}{2}\,J_{ij}\,
\nu_{i\sigma}\,\nu_{j\overline{\sigma}}\right)\,,
\label{eq:w10}
\end{equation}
where $J_{ij}$ expresses the sum in the bracket in (\ref{eq:w9}).
This is the spin-dependent version of an attractive density-density
interaction [12], which still expresses the complementary character of
antiferromagnetic spin-spin correlations and the intersite singlet pairing
(note that $\nu_{i\sigma}\,\nu_{j\overline{\sigma}}=
b_{i\sigma}^{\dagger}b_{j\overline{\sigma}}^{\dagger}b_{j\overline{\sigma}}
b_{i\sigma}$).

The essence of the above argument is as follows. The inclusion of a bond
distortion associated with the particle density-density correlation
($<\nu_{i}\nu_{j}>\neq 0$) results in an effective t-J model of the type
(\ref{eq:w10}). The bond-distortion contribution can be large as the
ratio of the bare intersite Coulomb interaction $(W_{ij})$ to
$\gamma_{1}R_{ij}^{2}$ can be large and therefore, together
with the kinetic-exchange contribution $(\sim t_{ij}^{2})$,
can overcome the intersite repulsion $\sim K_{ij}$ between the particles
with the opposite spins. In this manner, the t-J of model looks as a
model of real-space pairing in a single narrow band, in which the Cooper pairs
are formed as bound states in an attractive potential. In the absence of the
bond distortion, the Cooper-pair and the condensed states are formed
purely by the electronic correlations induced by the repulsive
Coulomb interactions.

\section{Hybrid singlet $d-p$ vs. $d-d$ pairing: A brief comment}

The magnetism of strongly correlated electrons started with pioneering
introduction of Hubbard model by Anderson [13,14] in the context of
antiferromagnetic ordering of transition-metal oxides. In this treatment
the role of the filled $2p$ states due to oxygen (O$^{2-}$) is passive, i.e.
to mediate the effect of $d-d$ interaction between the magnetic ions.
The role of oxygen is not only passive in e.g. metal-insulator transition
[15]. The question is what happens if the state is metallic and the $p$ bands
becomes partially filled and the $p-d$ hybridization starts playing an active
role in the dynamics in the metallic phase. The passive role of the $p$
electrons has been assumed by Zhang and Rice [16] in their intuitive
justification of the t-J model by assuming that the Kondo-like (Zhang-Rice)
singlet is so tightly bound that the $p-d$ antiferromagnetic correlations
lead to effective $d-d$ interactions. The situation seems very likely for the
cuprates (see however e.g. [17]). However, it is certainly not the case
for the heavy fermions, where the hybridization between the strongly correlated
$4f$ states (of e.g. Ce$^{3+}$ ion) and the $5d-6s$ valence states destabilizes
the $4f$ localized moments, in effect leads to the heavy-fermion behavior,
and can become the source of the hybrid pairing at the same time
[8,10]. Parenthetically,
the situation may be different in heavy-fermion uranium compounds such as
UPt$_{3}$,
where the U$^{4+}$ (approximately $5f^{2}$) configuration may lead,
in conjunction with the Hund's rule coupling among $5f$ electrons,
to the triplet pairing. This type of
situation will be considered in the next section.

\section{Real-space pairing in the doubly degenerate band}

\subsection{The physical situation and the model}

As we have seen in the two foregoing Sections, the real space pairing
in the orbitally nondegenerate case leads, even in the two-band situation,
to the singlet correlations and the corresponding bound Cooper pairs [19].
We discuss now the situation in the orbitally degenerate Hubbard model
and include the Hund's rule coupling directly, as well as highlight its
role in the strong-correlation limit, where both ferro- and antiferro-magnetic
kinetic exchange interactions appear at a proper band filling. The purpose
of this discussion is to show that the real-space spin-triplet pairing is
theoretically as feasible as is the singlet pairing in the single-band case.

The starting Hamiltonian for such a two-band case has the following form

\begin{equation}
H\,=\,\sum_{ijll^{\prime}\sigma} {'}\, t_{ij}^{ ll^{\prime}}\,
a_{il\sigma}^{\dagger}\, a_{jl^{\prime}\sigma}\,+\,
U\,\sum_{il}\,n_{il\uparrow}\, n_{il\downarrow}
\label{eq:w11}
\end{equation}
$$
\sum_{il\neq l^{\prime}}\,\left[ \left( U^{\prime}-\frac{1}{2}\, J_{H}\right)
\,n_{il}n_{il^{\prime}}\,-\, J_{H}{\bf S}_{il}\cdot{\bf S}_{il^{\prime}}\,+\,
J_{H} a_{il\uparrow}^{\dagger}\, a_{il\downarrow}^{\dagger}
a_{il^{\prime}\downarrow}\, a_{il^{\prime}\uparrow}\right]\,,
$$
where $l,l^{\prime}=1,2$ are the orbital indices, the first and the second terms
are respectively the hopping and the Hubbard parts, whereas the last three
terms describe the interorbital Coulomb repulsion, the ferromagnetic Hund's rule
and the pair-electron-hopping terms, respectively (both have the same amplitude
$J_{H}$).
Without a loss of generality, one may assume that $U^{\prime}=U-2J_{H}$
which holds at least for $3de_{g}$ orbitals. To simplify the discussion, we
assume that the hopping is diagonal and the same, i.e. $ t_{ij}^{ ll^{\prime}}
=t_{ij}\delta_{ll^{\prime}}$ to avoid the complications associated with the
interorbital hopping part (with $l\neq l^{\prime}$) and the corresponding
exchange.

The model in the strong-correlation limit $U\gg J_{H}\gg\vert t_{ij}\vert$ leads
to ferromagnetic Mott insulator (with possible orbital ordering) for
quarter-filled band ($n=1$) and to the ordinary antiferromagnetic
Mott insulator at half filling ($n=2$) [20-22]. The reason for
ferromagnetism near $n\sim 1$ is due to the fact, that the dominant virtual
hopping process between two neighboring sites takes place to the local
spin-triplet state in the intermediate state. This intermediate on-site
spin-triplet state does not appear in an orbitally nondegenerate model. On
the other hand, for the case of two electron per atom ($n=2$) only the
virtual-hopping processes leading to the spin-singlet pair configuration
in the intermediate state are allowed, so the kinetic exchange interaction
is antiferromagnetic, in a direct analogy to that in a single-band case.
This gradual transition from paramagnetism through ferromagnetism to
antiferromagnetism has been observed in the series of isostructural
compounds FeS$_{2}$-CoS$_{2}$-NiS$_{2}$ and their mixtures, as then a doubly
degenerate band of $e_{g}$ symmetry is being filled when we go from FeS$_{2}$
($n=0$), through CoS$_{2}$ ($n=1$), to NiS$_{2}$ ($n=2$). At an intermediate
composition $1<n<2$ a spin-glass-like state appears and is caused by the
competing ferro- and antiferro-magnetic interactions [23].

\subsection{Spin-triplet paired states}

The model of the real-space pairing is built in a direct correspondence to the
theory of kinetic exchange. We consider the case of identical orbitals,
for which the nearest hopping integrals are the same $t\equiv t_{<ij>}$.
The relevant virtual hopping processes are depicted in Fig. 2. The effective
t-J-J$_{H}$ model takes the form for $n\leq 1$ [24]

$$
P_{0}\tilde{H}P_{0}\,=\,
\sum_{ijl\sigma}{'}\, t_{ij}\,
b_{il\sigma}^{\dagger}\, b_{jl\sigma}\,-\,
\frac{4}{U-3J}\,
\sum_{\stackrel{i\neq j\neq r}{m=-1,0,1}}\, t_{ij}t_{jr}
B_{ijm}^{\dagger}\, B_{ijn}
$$
$$
-\,\frac{4}{U-J}\,\sum_{i\neq j\neq r}\, t_{ij}t_{jr}
C_{ij0}^{\dagger}\, C_{rj0}\,-\,\frac{2}{U}\,
\sum_{\stackrel{i\neq j\neq r}{m=\pm 1}}\, t_{ij}t_{jr}
C_{ijm}^{\dagger}\, C_{jrm}
$$
\begin{equation}
+\,\frac{4J}{U^{2}}\,
\sum_{\stackrel{i\neq j\neq r}{\overline{m}=-m=\pm 1}}\, t_{ij}t_{jr}
C_{ijm}^{\dagger}\, C_{rj\overline{m}}\,,
\label{eq:w12}
\end{equation}
where
$b_{i1\sigma}^{\dagger}\equiv a_{i1\sigma}^{\dagger}(1-n_{i1\overline{\sigma}})
(1-n_{i2\overline{\sigma}})$, etc. The pairing operators with $-1,0,+1$ are the
projected interorbital pairing operators for the spin-triplet states with
$S^{z}=1,-1$, and $0$, respectively, i.e.
\begin{equation}
\begin{array}{lcl}
B_{ij1}^{\dagger}\,\equiv\, b_{i1\uparrow}^{\dagger} \,
b_{j2\uparrow}^{\dagger} & \mbox{for}
& m\,=\, S^{z}\,=\, 1 \\
B_{ij-1}^{\dagger}\,\equiv\, b_{i1\downarrow}^{\dagger}
\, b_{j2\downarrow}^{\dagger}& \mbox{for}
& m\,=\, S^{z}\,=\, -1 \\
B_{ij0}^{\dagger}\,\equiv\,\frac{1}{\sqrt{2}}\,
\left( b_{i1\uparrow}^{\dagger}\, b_{j2\downarrow}^{\dagger}\,-\,
b_{i1\downarrow}^{\dagger}\, b_{j2\uparrow}^{\dagger}\right) & \mbox{for}
& m\,=\, S^{z}\,=\, 0\,.
\end{array}
\label{eq:w13}
\end{equation}

The operators $C$ are the spin-singlet pairing operators which we call
the orbital pairing operators. This because they are defined as follows
\begin{equation}
\begin{array}{lll}
C_{ij1}^{\dagger}\,=\,\frac{1}{\sqrt{2}}\,\left(
b_{i1\uparrow}^{\dagger} \, b_{j1\downarrow}^{\dagger}\,-\,
b_{i1\downarrow}^{\dagger} \, b_{j1\uparrow}^{\dagger}\right)
& \phantom{aa} & m\,\equiv\, L^{z}\,=\, 1\,, \\
C_{ij\overline{1}}^{\dagger}\,=\,\frac{1}{\sqrt{2}}\,\left(
b_{i2\uparrow}^{\dagger} \, b_{j2\downarrow}^{\dagger}\,-\,
b_{i2\downarrow}^{\dagger} \, b_{j2\uparrow}^{\dagger}\right)
& \phantom{aa} & m\,\equiv\, L^{z}\,=\, -1\,, \\
C_{ij0}^{\dagger}\,\equiv\,\frac{1}{2}\,
\left( b_{i1\uparrow}^{\dagger}\, b_{j2\downarrow}^{\dagger}\,+\,
b_{i2\uparrow}^{\dagger}\, b_{j1\downarrow}^{\dagger}\,-\,
b_{i1\downarrow}^{\dagger} \, b_{j2\uparrow}^{\dagger}\,-\,
b_{i2\downarrow}^{\dagger} \, b_{j1\uparrow}^{\dagger}
\right)
& \phantom{aa} & m\,\equiv\, L^{z}\,=\, 0\,.
\end{array}
\label{eq:w14}
\end{equation}

The pairing operators $C_{ijm}^{\dagger}$ are both intraorbital
($m=\pm 1$) and interorbital ($m=0$) character. We see that the spin-triplet
term is dominant for $n\leq1$.
One should also note, that the intersite projected
pairing operators (\ref{eq:w13}) replace their intraatomic correspondant
[25,26], which appear in the limit of low and intermediate correlations
and are related then directly to the intraatomic ferromagnetic Hund's rule
coupling. Therefore, our work shows that the Hund's rule coupling can lead
not only to ferromagnetism (e.g. when the Stoner criterion is fulfilled),
but also to spin-triplet superconductivity in both moderate- and
strong-correlation regimes. It is tempting to propose that the spin-triplet
superconductivity may appear i pure samples of e.g. CoS$_{2}$ under pressure,
when ferromagnetism is suppressed. The question of a coexistence of the
spin-triplet superconductivity and ferromagnetism is a separate issue.

\begin{figure}[htb]
\centerline{\includegraphics[width=0.65\textwidth]{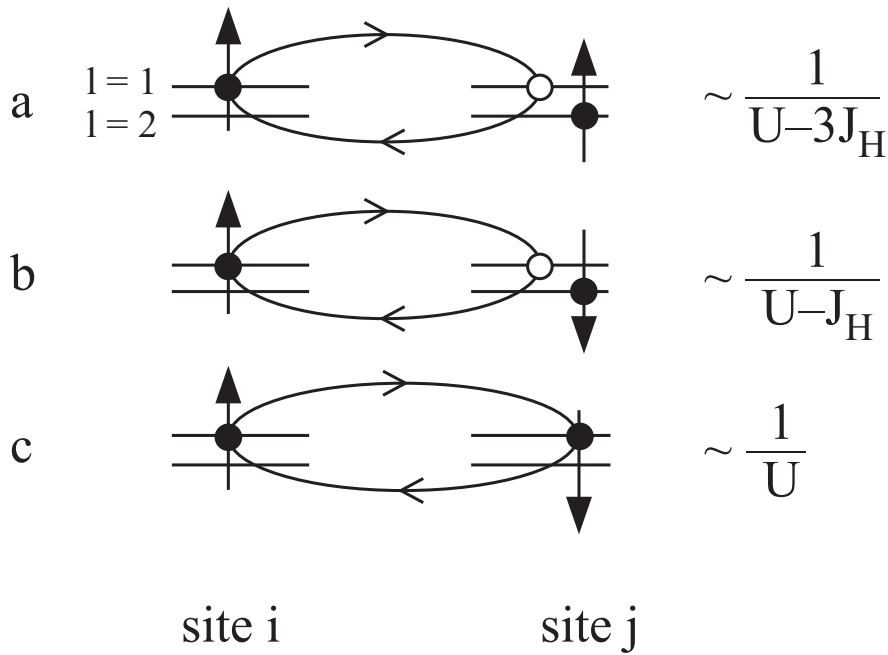}}
\caption{Possible virtual hopping processes leading to the three pairing
terms shown in the text (the corresponding denominators in Eq. (\ref{eq:w12})
are written on the right).}
\label{fig:f2}
\end{figure}

\section{Phase diagram of spin-triplet paired states}

Finally, we provide an overall phase diagram of the spin-triplet phases
and leave the details to a separate paper [27]. We consider a two-dimensional
square lattice case
with the isotropic gap ($\Delta_{m}=\Delta$). The superconducting
order parameter is chosen in the form
\begin{equation}
\Delta_{m{\bf k}}\,=\,\Delta_{f}\,\left(\cos k_{x}\,+\,
e^{i\alpha}\,\cos k_{y}\right)\,.
\label{eq:w15}
\end{equation}

In the slave-boson language, developed first for t-J model [28] this
fermionic gap amplitude $\Delta_{f}$ is multiplied additionally by the
slave boson
occupancy $\Delta_{b}=<b_{i}^{\dagger}b_{j}>\approx b^{2}$, and the
superconducting state is achieved when both the fermionic parameter
$\Delta_{f}\sim <B_{ijm}^{\dagger}>$ and the bosonic amplitude $\Delta_{b}$
are simultaneously different from zero. The solutions then are of extended
$s$-wave type (if $\alpha=0$), $s+id$ type (when $\alpha=\pi /2$), and of $d$
type (if $\alpha=\pi$). In general, when $\alpha\neq (0,\pi /2, \pi)$ the state
is called $s-d$ mixed state. The phase diagram representing the stable phases
in the ground state is shown in Fig. 3 on the plane $\delta\equiv 1-n$
versus $J_{H}/U$.
The other parameter is $\vert t\vert /U=0.2$. For $\delta =0$ the
system is a ferromagnetic insulator, so the paired states evolve from a {\em
ferromagnetic Mott insulator\/}. Note the presence of a usual "dome" of $d$-wave
superconductivity, here located in the interval $0.04\leq \delta\leq 0.13$.
In the regime of low electron concentration $\delta\geq 0.20$ an ordinary
gapless ($s$-wave) type of pairing takes place. It should be underlined that
in order to determine stability of the phases we have to solve the system of equations for $\Delta_{f},\,\Delta_{b}$, and the chemical potential $\mu$,
for given phase factor $\alpha$. The details of this cumbersome analysis
will be discussed elsewhere [27].

\begin{figure}[htb]
\centerline{\includegraphics[width=0.65\textwidth]{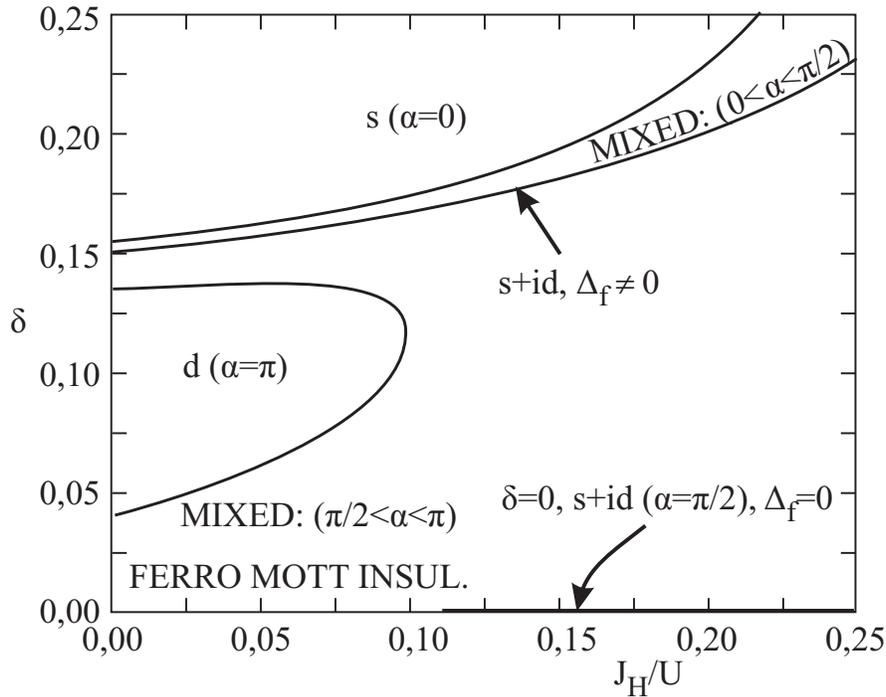}}
\caption{Paired ground states on the plane number of holes (per site)
in ferromagnetic Mott insulator versus the
relative strength of Hund's rule coupling.
For details see main text.}
\label{fig:f3}
\end{figure}

\section{Outlook}

In this brief overview we have first discussed the influence of the classical
bond
distortion in deriving an effective form of the spin-singlet pairing for
strongly correlated electrons in a nondegenerate band.
The inclusion of the lattice allows to draw a distinction between the
pairing achieved by the effective attractive interaction and that solely
by the correlations induced by the purely repulsive Coulomb interaction.
The role of hybrid
$2p-3d$ or $5d-4f$ correlations in deriving respectively the effective
spin-singlet pairing for high temperature and heavy-fermion systems is
noted in passing. The main message of our paper is though to indicate,
that an analogous type of pairing can take place for electrons in orbitally
degenerate and strongly correlated systems. Namely, we have discussed
the hole pairing in the doped {\em ferromagnetic Mott insulator\/}, with
the doping $\delta=1-n>0$. It would be interesting to see if such situation
is possible in one of the newly synthetized materials.
This new possibility at least extends the theoretical
applicability of the t-J model.

\section*{Acknowledgements}

This work is dedicated to Professor Ihor Stasyuk on the occasion
of his 70th Bithday. His papers on the Hubbard atomic representation
belongs to the pioneering works on strongly correlated systems.
I also acknowledge the Grant from the Ministry Of Science and
Higher Education of Poland, No. 1 P03B 029 001. The project is
being carried out under the auspices of the European COST P-16 Projects
{\em Emergent Behaviour of Correlated Matter\/}. The author is also
grateful to Drs. Andrzej Klejnberg and Robert Podsiad\l{}y for discussions
and technical help. This project is also partially supported by the Marie Curie TOK
Grant MTDK-CT-2004-517186 {\em Correlations in Complex Systems\/} (COCOS).


\begin{thebibliography}{99}

\bibitem{1}
A. F. Ruckenstein, P. Hirschfeld, and J. Appel, Phys. Rev. B
{\bf 36}, 857 (1987); N. M. Plakida, V. Yu. Yushankhai, and I. V. Stasyuk, Physica C {\bf 160}, 80 (1989).

\bibitem{2}
P. W. Anderson, in: {\em Frontiers and Borderlines in many-Particle Physics\/},
Eds. R. A. Broglia and J. R. Schrieffer, North-Holland, Amsterdam, 1988,
p. 1ff.

\bibitem{3}
J. Spa\l{}ek, Phys. Rev. B {\bf 38}, 208 (1988).

\bibitem{4}
J. Spa\l{}ek and P. Gopalan, J. Phys. (France) {\bf 50}, 2869 (1989).

\bibitem{5}
M. Ma\'ska et al., , Phys. Rev. Lett. {\bf 99}, 147006 (2007).

\bibitem{6}
S. S. Saxena et al., Nature {\bf 406}, 587 (2000).

\bibitem{7}
D. Aoki et al., Nature {\bf 413}, 613 (2001).

\bibitem{8}
A. Klejnberg and J. Spa\l{}ek, Phys. Rev. B {\bf 61}, 15542 (2000).

\bibitem{9}
J. Spa\l{}ek, P. Wr\'obel, and W. W\'ojcik, in {\em Ruthenate and
Rutheno-Cuprate Superconductors\/}, Eds. C. Noce et al., Springer Verlag,
Berlin 2002, vol. 603, pp. 60 - 75.

\bibitem{10}
For recent review see e.g.: J. Spa\l{}ek, Acta Phys. Polon. A {\bf 111},
409 (2007).

\bibitem{11}
J. Spa\l{}ek, to be submitted.

\bibitem{12}
R. Micnas, J. Ranninger, and S. Robaszkiewicz, Rev. Mod. Phys. {\bf 62},
113 (1990).

\bibitem{13}
P. W. Anderson, Phys. Rev. B {\bf 115}, 2 (1959).

\bibitem{14}
P. W. Anderson, in: {\em Solid State Physics\/},
Eds. F. Seitz and D. Turbull, Academic Press, New York, 1963,
vol. 14, p. 99ff.

\bibitem{15}
J. Zaanen, G. A. Sawatzky, and J. W. Allen,
Phys. Rev. Lett. {\bf 55}, 418 (1985); J. Magn. Magn. Mat. {\bf 55-57},
607 (1986).

\bibitem{16}
F. C. Zhang and T. M. Rice, Phys. Rev. B {\bf 37}, 3759 (1988).

\bibitem{17}
M. Eremin and A. Rigamonti, Phys. Rev. Lett. {\bf 88}, 037002 (2002).

\bibitem{18}
J. Spa\l{}ek, M. Ma\'ska, M. Mierzejewski, and J. Kaczmarczyk,
submitted for publication.

\bibitem{19}
K. Byczuk, J. Spa\l{}ek, and W. W\'ojcik, Phys. Rev. B {\bf 46}, 14134 (1992).

\bibitem{20}
K. I. Kugel and D. I. Khomskii, Sov. Phys.-JETP {\bf 37}, 725 (1973).

\bibitem{21}
M. Cyrot and C. Lyon-Caen, J. Phys. (France) {\bf 36}, 253 (1975).

\bibitem{22}
J. Spa\l{}ek and K. A. Chao, J. Phys. C: Sol. State Phys. {\bf 13},
5241 (1980).

\bibitem{23}
S. Ogawa, J. Appl. Phys. {\bf 50}, 2308 (1979).

\bibitem{24}
A. Klejnberg, Ph. D. Thesis, Jagiellonian University, Krak\'ow 2006,
unpublished.

\bibitem{25}
A. Klejnberg and J. Spa\l{}ek, J. Phys.: Condens. Matter {\bf 11},
6553 (1999).

\bibitem{26}
J. Spa\l{}ek, Phys. Rev. B {\bf 63}, 104513 (2001).

\bibitem{27}
A. Klejnberg and J. Spa\l{}ek, in preparation.

\bibitem{28}
Y. Suzumura, Y. Hasegawa, and H. Fukuyama, J. Phys. Soc. Jpn {\bf 57},
401, 2768 (1988); H. Fukuyama, Y. Hasegawa, and Y. Suzumura,
Physica C {\bf 153-155}, 1630 (1988).

\end{thebibliography}
\end{document}